\documentclass[eqsecnum,showkeys,showpacs,nofootinbib,aps,epsfig]{revtex4}
\renewcommand{\theequation}{\arabic{equation}}
\usepackage{graphicx}
\def\bea{\begin{eqnarray}}
\def\eea{\end{eqnarray}}

\newcommand{\nn}{\nonumber}
\newcommand{\na}{\nabla}
\def\beq{\begin{equation}}
\def\eeq{\end{equation}}

\begin{document}
\title{Hydrodynamics and global embeddings of Taub-NUT spacetime}
\author{Soon-Tae Hong}
\email{soonhong@ewha.ac.kr}
\author{Sung-Won Kim}
\email{sungwon@ewha.ac.kr}
\affiliation{Department of Science Education, Ewha Womans
University, Seoul 120-750 Korea}
%\date{January 22, 2003}%
\date{\today}%
\begin{abstract}
On Taub-NUT spacetime, we investigate hydrodynamic properties of
perfect fluid spiraling inward toward the spacetime along a
conical surface.  On the equatorial plane of the Taub-NUT
spacetime, we derive radial equations of motion with effective
potentials and the Euler equation for steady state axisymmetric
fluid. Higher dimensional global embeddings are constructed inside
and outside the event horizons of the Taub-NUT spacetime. We also
study the effective potentials
of particles on the Taub-NUT spacetime in terms of gravitational
magnetic monopole strength of the source, total energy and angular
momentum per unit rest mass of the particle.
\end{abstract}
\pacs{02.40.Ma, 04.20.Dw, 04.20.Jb, 04.70.Bw, 95.30.L}
\keywords{Taub-NUT spacetime, hydrodynamics, Euler equation,
global flat embedding} \maketitle

%%%%%%%%%%%%%%%%%%%%%%%%%%%%%%%%%%%%%%%%%%%%%%%%%%%%%%%%
\section{Introduction}
\setcounter{equation}{0}
\renewcommand{\theequation}{\arabic{section}.\arabic{equation}}
%%%%%%%%%%%%%%%%%%%%%%%%%%%%%%%%%%%%%%%%%%%%%%%%%%%%%%%%

For many decades, four dimensional solutions of the Einstein
equations have been extensively investigated in gravity community.
Recently solutions of the Einstein equations or the coupled
Einstein-matter equations in higher dimensions have attracted many
physical interests since the discovery of supergravity and
superstring theory. Taub-NUT metric~\cite{taub} is a local
analytic solution of the vacuum Einstein equations. When the
metric is expressed in Schwarzschild-like coordinates, we have
coordinate singularity occurring at certain values of the radial
coordinate where $g_{rr}$ becomes infinite and corresponds to
bifurcate Killing horizons. The Taub-NUT spacetime is involved in
many modern studies in general relativity.  Hawking has proposed
that the Euclidean Taub-NUT metric could give rise to the
gravitational analogue of the Yang-Mills
instanton~\cite{hawking77}.  In this case the Einstein equations
are fulfilled with zero cosmological constant and the manifold is
${\mathbf R}^{4}$ with a boundary which is a twisted three-sphere
possessing a distorted metric. The Kaluza-Klein monopole has been
obtained by embedding the Taub-NUT gravitational instanton into
five dimensional Kaluza-Klein theory~\cite{manton,atiyah}. The
non-Abelian target space duals of the Taub-NUT spacetime have been
also studied in terms of the local isometry group
SU(2)$\times$U(1)~\cite{hewson}.  The Taub-NUT spacetime has been
shown to be related with SU(2) calons through
T-duality~\cite{kraan98}. Carter has shown that the
Hamilton-Jacobi equation for the geodesics in the Taub-NUT metric
separates in certain coordinate systems~\cite{carter68}.  The
gravitomagnetic monopole source effects have been also studied in
the Taub-NUT spacetime~\cite{bini03}.  In the Kerr-Taub-NUT-de
Sitter metrics, separability of the Hamiltonian-Jacobi equation
has been studied in higher dimensions~\cite{chong05}. Recently, a
rotating Schwarzschild black hole has been studied to investigate
effective potentials for null and timelike geodesics of particles
and hydrodynamics associated with general relativistic Euler
equation for the steady state axisymmetric fluid~\cite{hong05}.

A familiar feature of exact solutions to the field equations of
general relativity is the presence of singularities. As novel ways
of removing the coordinate singularities, the higher dimensional
global flat embeddings of the black hole solutions are subjects of
great interest both to mathematicians and to physicists. It has
been well-known in differential geometry that four dimensional
Schwarzschild metric~\cite{sch} is not embedded in ${\mathbf
R}^{5}$~\cite{spivak75}. Recently, (5+1) dimensional global
embedding Minkowski space (GEMS) structure for the Schwarzschild
black hole has been obtained~\cite{deser97} to investigate a
thermal Hawking effect on a curved manifold~\cite{hawk75}
associated with an Unruh effect~\cite{unr} in these higher
dimensional spacetime.  In (3+1) dimensions, the global flat
embeddings inside and outside of event horizons of Schwarzschild
and Reissner-Nordstr\"{o}m black holes, have been constructed and
on these overall patches of the curved manifolds four
accelerations and Hawking temperatures have been evaluated by
introducing relevant Killing vectors~\cite{honggrg04}. Recently,
the GEMS scheme has been applied to stationary motions in
spherically symmetric spacetime~\cite{chen04}, and the
Banados-Teitelboim-Zanelli black hole~\cite{btz} has been embedded
in (3+2) dimensions to investigate the SO(3,2) global and Sp(2)
local symmetries~\cite{hongplb04}.

In this paper, exploiting the fact that the coordinates $t$ and
$\phi$ are cyclic in the Taub-NUT spacetime we find the timelike
Killing field and the axial Killing field, to which we can obtain
the conserved energy and the angular momentum per unit rest mass
for geodesics.  In the Taub-NUT spacetime we then investigate
hydrodynamic properties of the perfect fluid spiraling inward
toward the Taub-NUT spacetime along a conical surface. On the
equatorial plane of the Taub-NUT spacetime, we derive the radial
equations of motion with the effective potential.  We also study
the effective potentials of particles on the
Taub-NUT manifold in terms of gravitational magnetic monopole
strength of the source, total energy and angular momentum per unit
rest mass of the particle. Finally, we construct the higher
dimensional global embeddings inside and outside the event
horizons of the Taub-NUT manifold.

This paper is organized as follows. In section II we introduce the
Taub-NUT spacetime to study its hydrodynamics, and in section III
we construct the global flat embeddings inside and outside of
event horizon of the Taub-NUT spacetime. Section IV includes
summary and discussions.

%%%%%%%%%%%%%%%%%%%%%%%%%%%%%%%%%%%%%%%%%%%%%%%%%%%%%%%%
\section{Hydrodynamics of Taub-NUT spacetime}
\setcounter{equation}{0}
\renewcommand{\theequation}{\arabic{section}.\arabic{equation}}
%%%%%%%%%%%%%%%%%%%%%%%%%%%%%%%%%%%%%%%%%%%%%%%%%%%%%%%%

We start with the Taub-NUT 4-metric given by \beq
ds^{2}=-\frac{\Delta}{\Sigma}(dt+2l\cos\theta d\phi)^{2}
+\frac{\Sigma}{\Delta}dr^2+\Sigma(d\theta^{2}+\sin^{2}\theta
d\phi^2)\label{taubmetric}\eeq where, in the units $G=c=1$, \beq
\Delta=r^{2}-2Mr-l^{2},~~~ \Sigma=r^{2}+l^{2}, \label{delsig}\eeq
with the parameters $M$ and $l$ being associated with the mass and
gravitational magnetic monopole strength of the source. Defining
$r_{\pm}$ as \beq r_{\pm}=M\pm(M^{2}+l^{2})^{1/2}\label{rpm}\eeq
we can rewrite $\Delta$ as \beq \Delta=(r-r_{+})(r-r_{-}). \eeq
Here note that $r_{+}>l$ is an event horizon but $r_{-}$ is not
since $r_{-}$ is negative.

The four velocity is given by \beq u^{a}=\frac{d x^{a}}{d
\tau},\label{ua}\eeq where we can choose $\tau$ to be the proper
time (affine parameter) for timelike (null) geodesics. From the
equation of motion of a test particle in the Taub-NUT spacetime,
the particle initially at rest at infinity spiral inward toward
the spacetime along a conical surface of constant
$\theta=\theta_{\infty}$ where $\theta_{\infty}$ is the polar
angle at infinity.  For a fluid which is at rest at infinity and
approaches supersonically to the spacetime, one may take the
approximation to simplify the hydrodynamical equations \beq
u^{\theta}=\frac{d\theta}{d\tau}\approx 0. \eeq

As in the Schwarzschild black hole since the coordinates $t$ and
$\phi$ are cyclic we have the timelike Killing field $\xi^{a}$ and
the axial Killing field $\psi^{a}$. Corresponding to the Killing
fields $\xi^{a}$ and $\psi^{a}$ we can then find the conserved
energy $E$ and the angular momentum $L$ per unit rest mass for
geodesics given as follows \bea E&=&-g_{ab}\xi^{a}u^{b}\nn\\&=&
\frac{(r-r_{+})(r-r_{-})}{r^{2}+l^{2}}(u^{t}+2lu^{\phi}),\nn\\
L&=&g_{ab}\psi^{a}u^{b}\nn\\
&=&-\frac{2l(r-r_{+})(r-r_{-})}{r^{2}+l^{2}}~u^{t}
+\left((r^{2}+l^{2})\sin^{2}\theta-\frac{4l^{2}(r-r_{+})(r-r_{-})}
{r^{2}+l^{2}}\cos^{2}\theta\right)~u^{\phi}, \label{eandl} \eea
where $u^{a}$ are four velocity of the locally nonrotating
observers defined by (\ref{ua}). Moreover, we can introduce a new
conserved parameter $\kappa$ defined as \beq
\kappa=-g_{ab}u^{a}u^{b}\label{kappaeq}\eeq whose values are given
by $\kappa=1$ for timelike geodesics and $\kappa=0$ for null
geodesics.

In the case of geodesics on the equatorial plane $\theta=\pi/2$,
$u^{t}$ and $u^{\phi}$ are given in terms of $E$ and $L$ as
follows \bea
u^{t}&=&\frac{r^{2}+l^{2}}{(r^{2}+l^{2})^{2}+4l^{2}(r-r_{+})(r-r_{-})}
\left(\frac{(r^{2}+l^{2})^{2}}{(r-r_{+})(r-r_{-})}E-2lL\right)\nn\\
u^{\phi}&=&\frac{r^{2}+l^{2}}{(r^{2}+l^{2})^{2}+4l^{2}(r-r_{+})(r-r_{-})}
\left(L+2lE\right), \label{tphidot}\eea which are substituted into
(\ref{kappaeq}) to yield the radial equation for the particle on
the equatorial plane \beq
\frac{1}{2}E^{2}=\frac{1}{2}u^{r}u^{r}+V(r,E,L),\label{eomr}\eeq
with the effective potential \beq
V=\frac{(r-r_{+})(r-r_{-})}{2(r^{2}+l^{2})}\kappa
+\frac{(r-r_{+})(r-r_{-})}{2[(r^{2}+l^{2})^{2}+4l^{2}(r-r_{+})(r-r_{-})]}(L+2lE)^{2}.
\label{effpot}\eeq

For the null geodesics with $\kappa=0$, the effective potential
$V(x,y)$ for the particles with the total energy per unit rest
mass $E=0.1$ and angular momentum per unit rest mass $L=0.2M$ is
shown in the first graph in Fig. 1 where $x$ and $y$ denote the
dimensionless variables $x=r/M$ and $y=l/M$, respectively.  We
observe here that there exist many peaks on the potential surface.
In the third graph, we plot the surface
$z=(x^{2}+y^{2})^{2}+4y^{2}(x^{2}-2x-y^{2})$ and the plane $z=0$,
which shows the intersection curve corresponding to the vanishing
values of the denominator of the second term in (\ref{effpot}).
Along this curve we have peaks of the potential surface in the
first graph.  In the second graph in Fig. 1, $V(x,y)$ is shown for
the timelike geodesics of the particles with $E=0.1$ and $L=0.2M$.
In this case there exist peaks similar to those in the null
geodesics. The potential surface is deformed due to the first term
proportional to $\kappa$ in (\ref{effpot}), which however does not
affect the peak curve in the third graph since the denominator of
this term is positive definite.

The fundamental equations of relativistic fluid dynamics can be
obtained from the conservation of particle number and
energy-momentum fluxes.  In order to derive an equation for the
conservation of particle numbers one can use the particle flux
four vector $nu^{a}$~\cite{landau} \beq
\na_{a}(nu^{a})=\frac{1}{\sqrt{-g}}\na_{a}(\sqrt{-g}~nu^{a})=0,
\label{conteq} \eeq where $n$ is the proper number density of
particles measured in the rest frame of the fluid and $\na_{a}$ is
the covariant derivative in the Taub-NUT curved manifold and
$g={\rm det}~g_{ab}$.  For steady state axisymmetric flow, the
conservation of energy-momentum fluxes is similarly described by
the Einstein equation~\cite{shapiro} \beq
\na_{b}T_{a}^{b}=\frac{1}{\sqrt{-g}}\na_b(\sqrt{-g}~T_{a}^{b})=0,
\label{eineq} \eeq where the stress-energy tensor $T^{ab}$ for
perfect fluid is given by \beq T^{ab}=\rho
u^{a}u^{b}+(g^{ab}+u^{a}u^{b})P\eeq with $\rho$ and $P$ being the
proper internal energy density, including the rest mass energy,
and the isotropic gas pressure, respectively.  The Einstein
equation (\ref{eineq}) can be rewritten in another covariant form
\beq
u_{a}\na_{b}((P+\rho)u^{b})+(P+\rho)u^{b}\na_{b}u_{a}+\na_{a}P=0.
\label{eineq2} \eeq

On the other hand, from (\ref{taubmetric}), (\ref{delsig}) and
(\ref{rpm}) we can find the inverse of the Taub-NUT metric \bea
g^{tt}&=&-\frac{r^{2}+l^{2}}{(r-r_{+})(r-r_{-})}+\frac{4l^{2}(r^{2}+l^{2})}
{[(r^{2}+l^{2})^{2}+4l^{2}(r-r_{+})(r-r_{-})]\sin^{2}\theta},~~~
g^{rr}=\frac{(r-r_{+})(r-r_{-})}{r^{2}+l^{2}},~~~
g^{\theta\theta}=\frac{1}{r^{2}+l^{2}},\nn\\
g^{\phi\phi}&=&\frac{r^{2}+l^{2}}{[(r^{2}+l^{2})^{2}
+4l^{2}(r-r_{+})(r-r_{-})]\sin^{2}\theta},~~~
g^{t\phi}=-\frac{2l(r^{2}+l^{2})}{[(r^{2}+l^{2})^{2}
+4l^{2}(r-r_{+})(r-r_{-})]\sin^{2}\theta}. \label{inverse}\eea
Exploiting the inverse 4-metric $g^{ab}$ in (\ref{inverse}) and
acting the projection operator $g_{ab}+u_{a}u_{b}$ on the equation
(\ref{eineq2}) we can obtain the general relativistic Euler
equation on the direction perpendicular to the four velocity \beq
(P+\rho)u^{b}\na_{b}u_{a}+(g_{ab}+u_{a}u_{b})\na^{b}P=0.
\label{euler} \eeq After some algebra, from (\ref{euler}) we
obtain the radial component of the Euler equation for the steady
state axisymmetric fluid \bea & &
\frac{d}{dr}(u^{r}u^{r})+\frac{2r-r_{+}-r_{-}}{r^{2}+l^{2}}
-\frac{2(r-r_{+})(r-r_{-})r}{(r^{2}+l^{2})^{2}}
+\frac{4l^{2}B}{(r^{2}+l^{2})A}\nn\\
& &+\frac{4l^{2}B}{(r-r_{+})(r-r_{-})A}u^{r}u^{r}
+\frac{2}{P+\rho}\left(u^{r}u^{r}+\frac{(r-r_{+})(r-r_{-})}{r^{2}+l^{2}}\right)\frac{dP}{dr}=0,
\label{reuler}\eea where \bea A&=&(r^{2}+l^{2})^{2}
\sin^{2}\theta-4l^{2}(r-r_{+})(r-r_{-})\cos^{2}\theta\nn\\
B&=&\frac{(r^{2}+l^{2})(r-r_{+})(r-r_{-})[(2r-r_{+}-r_{-})(r^{2}+l^{2})-4r(r-r_{+})(r-r_{-})]}
{(r^{2}+l^{2})^{2}+4l^{2}(r-r_{+})(r-r_{-})}.\eea In the vanishing
$l$ limit, the result (\ref{reuler}) is reduced to that of the
(rotaing) Schwarzschild black hole~\cite{hong05}.

Multiplying (\ref{eineq2}) by $u^{a}$ we can project it on the
direction of the four velocity to obtain \beq
nu^{a}\left(\na_{a}\left(\frac{P+\rho}{n}\right)-\frac{1}{n}\na_{a}P\right)=0,
\label{nua}\eeq where the continuity equation (\ref{conteq}) has
been used.  The radial component of (\ref{nua}) yields \beq
\frac{d\rho}{dr}-\frac{P+\rho}{n}\frac{dn}{dr}=\frac{\Lambda-\Gamma}{u^{r}}.
\label{rrho} \eeq Here the energy loss $\Lambda$ and the energy
gain $\Gamma$ are introduced to set the decrease in the entropy of
inflowing gas equal to the difference $\Lambda-\Gamma$. Here it is
interesting to see that the result (\ref{rrho}) holds also in the
(rotating) Schwarzschild black hole~\cite{hong05}.

%%%%%%%%%%%%%%%%%%%%%%%%%%%%%%%%%%%%%%%%%%%%%%%%%%%%%%%%
\section{Global embeddings of Taub-NUT spacetime}
\setcounter{equation}{0}
\renewcommand{\theequation}{\arabic{section}.\arabic{equation}}
%%%%%%%%%%%%%%%%%%%%%%%%%%%%%%%%%%%%%%%%%%%%%%%%%%%%%%%%

After tedious algebra, for the Taub-NUT spacetime in the region
$r\ge r_{+}$ we can obtain the (6+5) global embedding Minkowski
space (GEMS) structure \beq
ds^{2}=-(dz^{0})^{2}+(dz^{1})^{2}-(dz^{2})^{2}+(dz^{3})^{2}+(dz^{4})^{2}+(dz^{5})^{2}+(dz^{6})^{2}
-(dz^{7})^{2}-(dz^{8})^{2}+(dz^{9})^{2}-(dz^{10})^{2} \label{gems}
\eeq with the coordinate transformations \bea
z^{0}&=&\kappa^{-1}\left(\frac{(r-r_{+})(r-r_{-})}{r^{2}+l^{2}}\right)^{1/2}
\cos\frac{\theta}{2}\sinh \kappa (t+2l\phi),\nn\\
z^{1}&=&\kappa^{-1}\left(\frac{(r-r_{+})(r-r_{-})}{r^{2}+l^{2}}\right)^{1/2}
\cos\frac{\theta}{2}\cosh \kappa (t+2l\phi),\nn\\
z^{2}&=&\kappa^{-1}\left(\frac{(r-r_{+})(r-r_{-})}{r^{2}+l^{2}}\right)^{1/2}
\sin\frac{\theta}{2}\sinh \kappa (t-2l\phi),\nn\\
z^{3}&=&\kappa^{-1}\left(\frac{(r-r_{+})(r-r_{-})}{r^{2}+l^{2}}\right)^{1/2}
\sin\frac{\theta}{2}\cosh \kappa (t-2l\phi),\nn\\
z^{4}&=&\left(\frac{(r^{2}+l^{2})^{2}+4l^{2}(r-r_{+})(r-r_{-})}
{r^{2}+l^{2}}\right)^{1/2}\sin\theta\cos\phi,\nn\\
z^{5}&=&\left(\frac{(r^{2}+l^{2})^{2}+4l^{2}(r-r_{+})(r-r_{-})}
{r^{2}+l^{2}}\right)^{1/2}\sin\theta\sin\phi,\nn\\
z^{6}&=&\left(\frac{(r^{2}+l^{2})^{2}+4l^{2}(r-r_{+})(r-r_{-})}
{r^{2}+l^{2}}\right)^{1/2}\cos\theta,\nn\\
z^{7}&=&\left(4l^{2}+\frac{1}{4}\kappa^{-2}\right)^{1/2}
\left(\frac{(r-r_{+})(r-r_{-})}{r^{2}+l^{2}}\right)^{1/2}\cos\theta,\nn\\
z^{8}&=&\left(4l^{2}+\frac{1}{4}\kappa^{-2}\right)^{1/2}
\left(\frac{(r-r_{+})(r-r_{-})}{r^{2}+l^{2}}\right)^{1/2}\sin\theta,\nn\\
z^{9}&=&\int dr\frac{\{p(r)[(r_{+}-r_{-})(r^{2}+l^{2})^{2}+(r_{+}^{2}+l^{2})s(r)]
+q(r)(r_{+}^{2}+l^{2})(r_{+}+r_{-})l^{2}\}^{1/2}}
{(r-r_{-})^{1/2}(r_{+}-r_{-})(r^{2}+l^{2})^{3/2}}\equiv f(r),\nn\\
z^{10}&=&\int
dr\left(\frac{p(r)(r_{+}^{2}+l^{2})(r_{+}+r_{-})l^{2}
+q(r)[(r_{+}-r_{-})(r^{2}+l^{2})^{2}+(r_{+}^{2}+l^{2})s(r)]}
{(r-r_{-})(r_{+}-r_{-})^{2}(r^{2}+l^{2})^{3}}\right.\nn\\
& &~~~~~~~~~~\left.+\frac{[r(r^{2}+l^{2})^{2}+2l^{2}(2r-r_{+}-r_{-})
(r^{2}+l^{2})-4l^{2}r(r-r_{+})(r-r_{-})]^{2}}
{[(r^{2}+l^{2})^{2}+4l^{2}(r-r_{+})(r-r_{-})](r^{2}+l^{2})^{3}}\right)^{1/2}\equiv g(r),
\label{gems11}
\eea
where $\kappa$ is given by
\beq
\kappa^{-2}=\frac{16}{3}\left[\left(\frac{r_{+}^{2}+l^{2}}{r_{+}-r_{-}}\right)^{2}+l^{2}\right],
\label{kappa}
\eeq
and the positive definite functions are given by
\bea
p(r)&=&(r_{+}-r_{-})r^{3}+r_{+}(r_{+}-r_{-})r^{2}+l^{2}r_{+}r_{-},\nn\\
q(r)&=&(2r_{+}^{2}+3l^{2})r_{-}r+l^{2}(r_{+}^{2}+r_{+}r_{-}+2l^{2}),\nn\\
s(r)&=&(r_{+}-r_{-})r^{2}+2(l^{2}-r_{+}r_{-})r. \eea Here note
that $\kappa$ in (\ref{kappa}) is not the surface gravity,
different from those in the embeddings in the static black holes
without the shift
functions~\cite{deser97,hongprd00,hongplb04,honggrg04}.

Next, in order to investigate the region $r\le r_{+}$ we rewrite
the Taub-NUT 4-metric as \beq
ds^{2}=\left(-\frac{\Delta}{\Sigma}\right)(dt+2l\cos\theta
d\phi)^{2} -\left(-\frac{\Sigma}{\Delta}\right)dr^2+\Sigma
(d\theta^{2}+\sin^{2}\theta d\phi^2) \eeq in terms of the positive
definite lapse function inside the event horizon $r_{+}$ \beq
-\frac{\Delta}{\Sigma}=\frac{(r_{+}-r)(r-r_{-})}{r^{2}+l^{2}} \eeq
to yield the (6+5) GEMS structure (\ref{gems}) with the coordinate
transformation \bea
z^{0}&=&\kappa^{-1}\left(\frac{(r_{+}-r)(r-r_{-})}{r^{2}+l^{2}}\right)^{1/2}
\cos\frac{\theta}{2}\cosh \kappa (t+2l\phi),\nn\\
z^{1}&=&\kappa^{-1}\left(\frac{(r_{+}-r)(r-r_{-})}{r^{2}+l^{2}}\right)^{1/2}
\cos\frac{\theta}{2}\sinh \kappa (t+2l\phi),\nn\\
z^{2}&=&\kappa^{-1}\left(\frac{(r_{+}-r)(r-r_{-})}{r^{2}+l^{2}}\right)^{1/2}
\sin\frac{\theta}{2}\cosh \kappa (t-2l\phi),\nn\\
z^{3}&=&\kappa^{-1}\left(\frac{(r_{+}-r)(r-r_{-})}{r^{2}+l^{2}}\right)^{1/2}
\sin\frac{\theta}{2}\sinh \kappa (t-2l\phi),\nn\\
z^{7}&=&\left(4l^{2}+\frac{1}{4}\kappa^{-2}\right)^{1/2}
\left(\frac{(r_{+}-r)(r-r_{-})}{r^{2}+l^{2}}\right)^{1/2}\cos\theta,\nn\\
z^{8}&=&\left(4l^{2}+\frac{1}{4}\kappa^{-2}\right)^{1/2}
\left(\frac{(r_{+}-r)(r-r_{-})}{r^{2}+l^{2}}\right)^{1/2}\sin\theta,\nn\\
z^{9}&=&f(r),\nn\\
z^{10}&=&g(r), \eea where $(z^{4},z^{5},z^{6})$, $f(r)$ and $g(r)$
are the same as those in (\ref{gems11}).

%%%%%%%%%%%%%%%%%%%%%%%%%%%%%%%%%%%%%%%%%%%%%%%%%%%%%%%%
\section{Conclusions}
\setcounter{equation}{0}
\renewcommand{\theequation}{\arabic{section}.\arabic{equation}}
%%%%%%%%%%%%%%%%%%%%%%%%%%%%%%%%%%%%%%%%%%%%%%%%%%%%%%%%

In conclusion, we studied the Taub-NUT spacetime to investigate
hydrodynamic properties of the perfect fluid whirling inward
toward the Taub-NUT spacetime along a conical surface. On the
Taub-NUT manifolds we then constructed the (6+5) dimensional
global embeddings inside and outside the event horizons of the
manifold. We also studied the effective potentials of
particles on the equatorial planes of the Taub-NUT
spacetime in terms of the gravitational magnetic monopole strength
of the source, total energy and angular momentum per unit rest
mass of the particle.

\acknowledgments The work of STH was supported by the Korea
Research Foundation Grant funded by the Korean Government (MOEHRD) (KRF-2006-331-C00071). SWK was supported by the Korea
Research Foundation Grant funded by the Korean Government (MOEHRD) (KRF-2006-312-C00498).


\begin{thebibliography}{99}
\bibitem{taub} A. Taub, Ann. Math. {\bf 53}, 472 (1951); E.T.
Newman, L. Tamburino, and T. Unti, J. Math. Phys. {\bf 4}, 915
(1963).
\bibitem{hawking77} S.W. Hawking, Phys. Lett. {\bf A 60}, 81
(1977).
\bibitem{manton} N.S. Manton, Phys. Lett. {\bf B 110}, 54 (1985).
\bibitem{atiyah} M.F. Atiyah and N. Hitchin, Phys. Lett. {\bf A 107}, 21 (1985).
\bibitem{hewson} S. Hewson, Class. Quantum Grav. {\bf 13}, 1739
(1996).
\bibitem{kraan98} T.C. Kraan and P. van Baal, Phys. Lett. {\bf B
428}, 268 (1998).
\bibitem{carter68} B. Carter, Comm. Math. Phys. {\bf 10}, 280
(1968).
\bibitem{bini03} D. Bini, C. Cherubini, M. de Mattia and R.T.
Jantzen, Gen. Rel. Grav. {\bf 35}, 2249 (2003); D. Bini, C.
Cherubini, R.T. Jantzen and B. Mashhoon, Class. Quantum Grav. {\bf
20}, 457 (2003); D. Bini, C. Cherubini, R.T. Jantzen and B.
Mashhoon, Phys. Rev. {\bf D 67}, 084013 (2003).
\bibitem{chong05} Z.W. Chong, G.W. Gibbons, H. L\"u and C.N. Pope,
Phys. Lett. {\bf B 609}, 124 (2005).
\bibitem{hong05} S.T. Hong and S.W. Kim, {\tt gr-qc/0503079}.
\bibitem{sch} K. Schwarzschild, Sitzber. Deut. Akad. Wiss. Berlin, KI.
Math.-Phys. Tech. pp. 189-196 (1916).
\bibitem{spivak75} M. Spivak,
{\it Differential Geometry} (Publish or Perish, Berkeley 1975) Vol
5, Chapter 11.
\bibitem{deser97} S. Deser and O. Levin, Class. Quant. Grav. {\bf 14}, L163 (1997);
S. Deser and O. Levin, Class. Quant. Grav. {\bf 15}, L85 (1998);
S. Deser and O. Levin, Phys. Rev. {\bf D 59}, 064004 (1999).
\bibitem{hawk75} S.W. Hawking, Comm. Math. Phys. {\bf 42}, 199 (1975);
J.D. Bekenstein, Phys. Rev. {\bf D 7}, 2333 (1973); R.M. Wald,
{\em Quantum Field Theory in Curved Spacetime and Black Hole
Thermodynamics} (The University of Chicago Press, Chicago 1994);
J.D. Brown, J. Creighton and R.B. Mann, Phys. Rev. {\bf D 50},
6394 (1994); R.M. Wald, Living Rev. Rel. {\bf 4}, 6 (2001).
\bibitem{unr} W.G. Unruh, Phys. Rev. {\bf D 14}, 870 (1976);
P.C.W. Davies, J. Phys. {\bf A 8}, 609 (1975).
\bibitem{honggrg04} S.T. Hong, Gen. Rel. Grav. {\bf 36}, 1919 (2004).
\bibitem{chen04} H.Z. Chen, Y. Tian and Y.H. Gao, JHEP {\bf 0410}, 011 (2004);
H.Z. Chen and Y. Tian, {\tt gr-qc/0410077}.
\bibitem{btz} M. Banados, C. Teitelboim, and J. Zanelli, Phys. Rev. Lett. {\bf 69}, 1849 (1992);
M. Banados, M. Henneaux, C. Teitelboim, and J. Zanelli, Phys. Rev.
{\bf D 48}, 1506 (1993).
\bibitem{hongplb04} S.T. Hong, Phys. Lett. {\bf B 578}, 187
(2004).
\bibitem{landau} L.D. Landau and E.M. Lifschitz, {\it Fluid
Mechanics} (Pergamon, Oxford 1977).
\bibitem{shapiro} S.L. Shapiro, Ap. J. {\bf 189}, 343 (1974).
\bibitem{hongprd00} S.T. Hong, Y.W. Kim and Y.J. Park, Phys. Rev.
{\bf D 62}, 024024 (2000); S.T. Hong, W.T. Kim, Y.W. Kim and Y.J.
Park, Phys. Rev. {\bf D 62}, 064021 (2000); S.T. Hong, W.T. Kim,
J.J. Oh and Y.J. Park, Phys. Rev. {\bf D 63}, 127502 (2001).
\end{thebibliography}
\end{document}